\documentclass[aps,prb,superscriptaddress,twocolumn]{revtex4}
\usepackage[dvips]{graphicx} \usepackage{amsmath}

\begin{document}


\title{Self-consistent study of electron confinement 
to metallic thin films on solid surfaces 
}


\author{E. Ogando}\email{eoa@we.lc.ehu.es}\affiliation{Elektrika eta
Elektronika Saila, Zientzia Fakultatea UPV-EHU 644 P.K., 48080 Bilbao,
Spain} \author{ N. Zabala}\affiliation{Elektrika eta Elektronika
Saila, Zientzia Fakultatea UPV-EHU 644 P.K., 48080 Bilbao, Spain}
\affiliation{Donostia International Physics Center (DIPC) and Centro
Mixto CSIC-UPV/EHU, Apto. 1072, 20080 Donostia, Spain}
\author{E.V. Chulkov}\affiliation{Donostia International Physics
Center (DIPC) and Centro Mixto CSIC-UPV/EHU, Apto. 1072, 20080
Donostia, Spain} \affiliation{Materialen Fisika Saila, Kimika
Fakultatea, UPV-EHU 1072 P.K., 20080 Donostia, Spain}
\author{M.J. Puska} \affiliation{Laboratory of Physics, Helsinki
University of Technology, P.O. Box 1100, FIN-02015 HUT, Finland }

\date{\today}

\begin{abstract}
We present a method for density-functional modeling
of metallic overlayers grown on metallic supports. It offers a tool to study nanostructures and combines 
the power of self-consistent pseudopotential calculations with 
the simplicity of a one-dimensional approach. The model is 
applied to Pb layers grown on the Cu(111) surface. 
More specifically, Pb is modeled as stabilized jellium and 
the Cu(111) substrate is represented by a one-dimensional pseudopotential
that reproduces experimental positions of both the Cu Fermi level 
and the energy gap of the band structure projected along
the (111) direction. The model is used to study the
quantum well states in the Pb overlayer. Their analysis
gives the strength of the electron confinement barriers at the 
interface and at the surface facing the vacuum. Our results
and analysis support the interpretation of the 
quantum well state spectra measured by the scanning tunneling
spectroscopy. 

\end{abstract}



\maketitle


\section{Introduction}

The ability to manufacture desired nanostructures on solid surfaces
by controlled growth or atomic manipulation techniques has increased
enormously during the last decades. \cite{ortega,chiang} At the same time, 
spectroscopic methods to study different physical properties and phenomena
of these structures have also experienced a huge development. 
In understanding and interpreting the ensuing experimental results
rich of quantum phenomena, accurate theoretical and computational 
modeling plays a vital role. 

A widely studied phenomenon is the growth of thin Pb films or
extended Pb islands on solids, for example, on Si or Cu
surfaces. The growth of Pb provides a laboratory to test the
the so-called quantum size effects (QSE) arising due to the electron 
confinement perpendicular to the surface. The confinement results in 
discrete energy levels, the so-called quantum well states (QWS's). 
With the increasing film thickness new QWS's become occupied producing
oscillations in the total energy, work function and other physical 
properties. The oscillations in the energy are the origin of the "magic" 
islands heights.
\cite{otero,hinch,budde,luh,yeh2,hupalo,mans,hupalo2,su,menzel,edu_slab,
gavioli,nagao1,nagao2}

The most important feature characterizing an overlayer or a
nanoisland is its height. However, measurements based
on different physical processes may provide different values.
The x-ray diffraction \cite{czoschke} can be used to directly determine 
the number of atomic monolayers (ML) in thin films. The scanning 
tunneling microscope (STM) \cite{otero} gives the thickness 
of finite nanoislands. In the helium atom scattering (HAS) the specular
returning point of He atoms is measured.\cite{floreano,hinch}
STM and HAS reflect the electron density profile of the 
surface. \cite{floreano,hinch} The scanning tunneling spectroscopy 
(STS) measures the QWS spectrum and the width of the quantum well 
confining the electrons is evaluated from it. \cite{otero2}
In addition to the overlayer-vacuum surface profile, the QWS spectrum 
is sensitive also to properties of the substrate-overlayer
interface. The information about the confining quantum well
is important not only for the determination of the the film
thickness but also for the understanding and controlling
the so-called "electronic growth mode".

In this work we model thin Pb films growing in the [111] direction
on the Cu(111) surface. 
We focus is on determining the surface and interface barriers of 
the confining quantum well and the ensuing QWS's. 
Pb/Cu(111) was recently studied by Otero et. al. \cite{otero2} 
They fitted the QWS spectra measured by STS by using the 
finite-potential-well model and unexpectedly found that 
actually the infinitely high potential barriers give a better fit than 
the more realistic finite
barriers. One of our main aims is to resolve this dilemma. There 
are also recent STS measurements of QWS's in the Pb layers
on Si(111).\cite{altfeder,su,su2} In these studies
the overlayers are thinner and the QWS energy range is smaller
than in the Pb/Cu(111) measurements by Otero et. al. \cite{otero2}
The effective thickness of the Pb overlayer is then obtained by
fitting the energy {\em differences} of the consecutive states. This
is not very accurate and there are contradictory results for
the thickness of the wetting layer, ranging from 1 ML \cite{yeh2,hupalo,mans}
to 3 ML. \cite{altfeder,hupalo2,su,su2} 

There are several theoretical works based on the density functional 
theory (DFT) devoted to study QSE in thin films. First-principles 
atomistic approaches \cite{wei,materzanini,feibelman,feibelman2,kiejna,boettger,boettger2,saalfrank,batra}
or jellium models \cite{schulte,sarria,fiolhais,wojciechowski,ciraci}
have been used. Usually, due to the high computational demand the 
substrate is not included, but the electronic structure 
is calculated for a slab describing 
the overlayer. The work by Hong et al. \cite{hong} for Pb/Si(111)
is one exception. On the other hand, there are
several simple analytic models for the confinement barriers,
\cite{altfeder,otero2,echenique} but {\it a priori}
assumptions about the barrier type, its position or quantum numbers 
of the measured states can lead to an erroneous interpretation of the 
experiments. The simplifications made in the modeling hinder
effectively the analysis of the experimental results.

In the present work we perform self-consistent electronic structure
calculations by modeling the Cu(111) substrate with a
one-dimensional (1D) pseudopotential and the Pb overlayer by stabilized
jellium. The model includes the effect of the substrate-film boundary
so that the penetration of the QWS into the substrate is realistically
described. Our self-consistent results allow us to also study
the above-mentioned simple analytical models and point out their 
deficiencies as well as the most important factors for the
proper description of the electron confinement. This knowledge
is specially important when using these models in analysing 
the STS results for completely covered substrates or for 
systems with wetting layers of unknown thicknesses.

In a previous publication \cite{edu_slab} we applied successfully
the 1D-pseudopotential - stabilized-jellium model to gain physical
insight into the "magic" heights of Pb islands on Cu(111) up to 
23 ML of Pb.\cite{otero} A pragmatic aim of the present paper
is to document the construction of the unscreened 1D-pseudopotential and
provide a simple parametrization which can be used in future studies, 
e.g. for different nanostructures on surfaces. 

The rest of the paper is organized as follows: in Sec. \ref{theory} we
report the practical steps for the construction of the Cu(111)
pseudopotential. Then, we describe analytical models for the confinement
barriers, whose reliability is analyzed by applying them to 
the results of the
self-consistent calculations. Section \ref{results} demonstrates the
resulting self-consistent electronic structures for the Pb/Cu(111)
system. The confinement barriers are analyzed and an improvement to the
theory is introduced in order to acquire accuracy in their
determination. Section \ref{conclusions} contains the conclusions of
the work. Atomic units (\emph{i.e.}, $\hbar=e^{2}=m=1$ and distances
measured in Bohr radius units $a_0=0.53$ \AA) will be used throughout
this work, unless otherwise specified.

\section{Theory \label{theory}}
\subsection{The model}
The present calculations are performed in the framework of 
the DFT \cite{jones} within the local density
approximation (LDA) \cite{perdew-wang,ceperley-alder} and jellium-type models.
Instead of finite Pb islands we consider infinitely-extended films
on the Cu surface. This is justified because in 
the experiments considered the characteristic lateral dimension of the Pb
islands is around $1000a_0$ [\onlinecite{otero,otero2}] so that
the lateral electron confinement effecs are vanishingly small. 
Moreover, we assume the perfect translational invariance, i.e.
a homogeneous free electron gas, along the surface ($xy$ plane). 
We use a jellium model for the Pb overlayer and for the Cu substrate we construct
a pseudopotential which varies only along the $z$ 
direction perpendicular to the surface. Then the Kohn-Sham equations 
have to be solved numerically only in the $z$ direction (1D problem). This
enables the calculation of electron wavefunctions extending deep
into the Cu substrate and the modeling of systems having tens of
Pb ML's.

The Kohn-Sham equations are discretized in a regular one-dimensional
point mesh and solved with the Rayleigh quotient multigrid method,
\cite{mandel,heiskanen} implemented in the real-space MIKA package
\cite{tuomas} for electronic structure calculations.  Hence,
single-particle wave-functions are taken to be of the form
\begin{equation}
\Phi({\bf r})=\psi_n(z) e^{i{\bf k}_{\parallel}\cdot {\bf
r}_{\parallel}},
\end{equation}
where $\psi_n(z)$ is the wavefunction in the direction perpendicular
to the surface, and plane waves are used for the surface parallel
directions. The eigenenergies are given by
\begin{equation}
\epsilon_{nk_{\parallel}}=\epsilon_n+\frac{k_{\parallel}^2}{2},
\end{equation}
where $\epsilon_n$ is the eigenvalue of the $n$-th perpendicular state
$\psi_n(z)$. The eigenvalue $\epsilon_n$, obtained self-consistently, is the
bottom of the n-th subband.  
 For finite and periodic systems in the $z$ direction Dirichlet  
and periodic boundary conditions are used, respectively.

The effective or screened potential of the Kohn-Sham \cite{jones}
equations in the $z$ direction is written as
\begin {equation} \label{poteq}
V_{\rm eff}(z)=\int\frac{n_-(z')-n_+(z')}{|z-z'|}dr'+ V_{\rm
xc}[n_-(z)]+V_{ps}(z),
\end {equation}
where the first term on the right-hand side is the Hartree term
$V_H(z)$, which includes the electron density $n\_(z)$ and the
neutralizing rigid positive charge density $n_+(z)$. The second term 
gives the LDA
exchange-correlation potential. The third term accounts for the
pseudopotential that improves the simple jellium scheme. For the
supported overlayer system $V_{ps}(z)$ has two contributions, one from
the Pb overlayer and the other from the Cu(111) substrate. The
free electron-like character of Pb at the Fermi level justifies the
use of the stabilized jellium or averaged pseudopotential
\cite{perdew,shore2} approach to model Pb. In practice, the jellium
model allows us to simulate any Pb overlayer thickness. \cite{edu_slab} 
The Pb contribution to $V_{ps}(z)$ stabilizing the electron gas 
at the density corresponding to $r_s=2.30a_0$ is a constant shift
$V_{stab}$ relative to the vacuum level and restricted in the 
region of the positive background charge.
The stabilized Pb provides a proper work 
function so that the spilling of the electron density into the vacuum 
is well described. It also gives a proper value for
the bottom of the valence electron band. This guarantees the correct 
Fermi wave length $\lambda_F$, which is of
crucial importance for the properties related to the electron
confinement.

For the Cu(111) substrate it is necessary to use a pseudopotential
which accounts correctly for the width of the energy gap and its position
with respect to the Fermi level. In this way, we obtain the
correct confinement potential also at the Cu(111)-Pb interface. To obtain
the 1D-pseudopotential we use the procedure described in the
following subsection. We start from a model potential and
build an unscreened pseudopotential for Cu(111) in two steps.

\subsection{Generation of the Cu(111) 1D-pseudopotential}
\subsubsection{Bulk pseudopotential ($1^{st}$ step)}
Chulkov {\it et al.} \cite{chulkov} proposed a fully
screened 1D-model-potential which varies only in the $z$ direction
perpendicular to the surface. The model potential is
successfully used to study, for example, the dielectric 
response function and lifetimes of excited states. 
\cite{chulkov,chulkov2,chulkov3} The
crucial point here is the proper description of the energy
band gap and work function. Moreover, the wave functions are
correctly described not only outside the substrate, but 
also inside it. This is an important ingredient in the
present application.

Because in this work  we cover the Cu(111) surface with several ML's of Pb
we skip the surface part of the 1D-pseudopotential. Nevertheless, 
it is also possible to build a pseudopotential which reproduces 
the surface and image states. \cite{edu_unpublished}
The bulk oscillating function of the 1D-model-potential
\cite{chulkov} is
\begin{equation}
\label{bulk}
V_{model}(z)=A_{10}+A_1\cos \left ( \frac{2 \pi}{d}z \right ),\qquad \
\frac{-d}{2}<z<\frac{d}{2}
\end{equation}
where $d=3.943a_0$ is the interlayer spacing in Cu in the [111] 
direction and $A_{10}$ and $A_1$ are fitting parameters. 
Using periodic boundary conditions
at $\pm d/2$ the Kohn-Sham equations are solved for the fixed
$V_{eff}(z)=V_{model}(z)$ potential. With the eigenfunctions obtained 
and with the experimental work function
we calculate the electron
density profile. Integrating over $z$ we obtain the mean density with
$r_s=2.55a_0$. This value is close to the experimental $r_s=2.67a_0$ for 4s electrons.

Once we have computed the density, it is straightforward to obtain the
corresponding $V_{xc}(z)$ potential. It is more
challenging to calculate the Hartree $V_{H}(z)$ term because in absence 
of vacuum we don't know the height of the dipole barrier, and therefore, where to fix the energy origin for the Hartree term inside the bulk. To solve the problem we fix provisionally
the zero of the Hartree potential at the boundaries of the periodic
cell. After adding a homogeneous neutralizing positive background of
$r_s=2.55a_0$, the Hartree potential can be evaluated. Finally, we
can calculate from Eq. (\ref{poteq}) the unscreened and periodic
pseudopotential $V_{ps}(z)$ by subtracting from the effective
potential the $V_H(z)$ and $V_{xc}(z)$ terms.

We have obtained, in this first step, the unscreened
pseudopotential for periodic bulk calculations. Performing a
self-consistent calculation for the bulk Cu(111) with this unscreened
pseudopotential $V_{ps}(z)$ and a positive background density
corresponding to $r_s=2.55a_0$, we recover the experimental
Fermi level with respect to the vacuum.

\subsubsection{Pseudopotential for a semi-infinite system ($2^{nd}$ step)}

The unscreened 1D-pseudopotential from the first step is suitable for bulk calculations. But it cannot be used in slab calculations because the 
zero of the $V_H(z)$ potential was arbitrarily chosen in the previous step. We
complete the pseudopotential for slab calculations in this second step.

We build a semi-infinite Cu(111) slab by repeating the
pseudopotential for one ML of Cu. The slab is thick enough to avoid 
interaction between surfaces and finite-size effects in determining the
band structure. Then, enough vacuum is added on both 
sides to annul boundary effects at the borders of the calculation 
volume. 
In the self-consistent calculation for this system the
electron density spills out of Cu(111) to the vacuum
giving rise to the dipole Coulomb barrier which places the
Fermi level (and the whole band structure including the energy gap)
at the incorrect position with respect to the vacuum level. 

To correct the work function we shift the potential by a constant 
inside the Cu slab. Here we define the Cu(111) edge to be one half ML 
above the last atom plane.
But the potential shift also changes the electron spilling
into the vacuum and the dipole barrier. Thus, we have to find
the potential shift iteratively so that the correct value
for the work function is recovered.
The value of the shift depends also on the possible smoothing 
(see below) of the positive background charge profile at the surface.

The 1D-pseudopotential vanishes suddenly at the Cu(111) surface,
producing a discontinuity in the total potential.
We have  examined different ways to obtain a pseudopotential without
the discontinuity. However, the final results obtained do not differ
from the results obtained with the scheme described above and
presented in this paper.

In summary, we have obtained a 1D-pseudopotential 
which with the positive background density of $r_s=2.55a_0$ reproduces 
correctly the experimental Cu(111) work function and the [111]-projected 
band structure including the band gap.
The method presented here for the Cu(111) pseudopotential generation
is extensible to other substrates as well.

\subsubsection{Analytic expression of the 1D-pseudopotential}
The unscreened pseudopotential obtained can be fitted using the form
of Eq. \ref{bulk} for the screeneed model-potential.
The result is, $A_{10}=-1.89\ eV$ and $A_1=5.01\ eV$. This
pseudopotential is obtained with the positive charge background density
corresponding to $r_{s}=2.55a_0$ and a surface profile of the positive
charge given by a Fermi-like distribution
\begin{equation}\label{psd}
\rho(z)=\frac{\rho_{_{Cu}}}{e^{\frac{z-D_0}{\Delta z}}+1} .
\end{equation}
Above, $D_0$ is the surface edge position and $\Delta z=0.09a_0$ accounts
for the smoothing at the edge. The smoothing is used for numerical
reasons. 

\subsection {Phase accumulation model for confinement barriers}\label{analy_for}

We use the onedimesional-pseudopotential described above to simulate the Pb/Cu(111)
system. In order to rationalize the results in a quantitative way,
we employ simple analytical models reducing the information about 
the confinement barriers in few parameters. In particular, we are
going to use the so-called phase accumulation model.\cite{echenique} 
In this model a QWS in a confining potential fullfils the equation
\begin{eqnarray} \label{eche}
2 k_z D+\phi_{Cu(111)-Pb}+\phi_{Pb-vac}=2\pi (n-1),
\end{eqnarray}
where $k_z$ is the wavevector corresponding to the QWS kinetic energy,
$D$ is the width of the potential well to model the Pb film, $\phi_{Pb-vac}$ and
$\phi_{Cu(111)-Pb}$ are the phases of the eigenfunction accumulated at
the vacuum surface and at the interface, respectively, and  
$n=1,2,3,\ldots$ is the quantum number of the QWS. 
For an infinitely deep square potential well, the phase 
accumulated on each surface is $-\pi$.
For softer potential barriers the value increases. 

The phase shifts, usually calculated for model potentials given 
analytically, incorporate the physical features of the QWS's 
in a simple manner. An equivalent but more intuitive way is to apply the 
infinitely deep square potential well so that 
$\phi_{Pb-vac}=\phi_{Pb-Cu}=-\pi$, 
but using an effective width of $D'=D+\delta_{Cu(111)-Pb}+\delta_{Pb-vac}$.
Then Eq. (\ref{eche}) gives
\begin {equation}\label{eche2}
\delta_{Cu(111)-Pb}+\delta_{Pb-vac}=\frac{\pi n}{k_z}-D.
\end {equation}
Here, $\delta_{Cu(111)-Pb}$ and $\delta_{Pb-vac}$ arise from 
the wavefunction penetration into the Cu(111) substrate and the spill out 
into the vacuum, respectively. The idea is that $D'$ should give
a reliable estimate of the actual thickness of the overlayer.
Thus, as the phase shifts, also the effective well width depends on the QWS 
eigenenergy $\varepsilon_n$. It has been shown that the energy spectrum 
is very sensitive to the positions of the barriers and relatively insensitive 
to the barrier height.\cite{otero2} Therefore, a mean surface shift $\delta_0$ can 
reproduce quite accurately the spectrum. According to Eqs. (\ref{eche}) 
and (\ref{eche2}) the surface position shifts and phase shifts
are related by
\begin{equation}
\delta=\frac{\phi+\pi}{2k_z}.
\end{equation}

We consider below two analytical expressions for the surface-vacuum phase shift,
the finite-potential-step and the image-potential models.
The former gives the energy-dependent phase shift \cite{otero2,smith}
\begin{equation}
\phi_{Pb-vac}=2\arctan \left ( -\frac{k_{vac}}{k_z} \right )
\end{equation}
where $k_{vac}=\sqrt{2(-\varepsilon)}$ with the energy eigenvalue
$\varepsilon$ measured with respect to the vacuum level. 
This model has been used for the Cu-Pb interface barrier, 
too.\cite{otero2}
The image-potential model gives the phase shift \cite{echenique2}
\begin{equation} \label{imag_p}
\phi_{Pb-vac}=  \pi \,\sqrt{\frac{3.4 \rm{\ eV}}{-\varepsilon}}.
\end{equation}

The phase shift corresponding to the QWS wavefunction penetrating into
the Cu(111) depends on the position of the QWS energy eigenvalue
relative to the energy band gap. For example, the empiric formula
\cite{echenique2}
\begin{equation}\label{empiric}
\phi_{Cu(111)-Pb}= 2 \arcsin \sqrt{ \frac{\varepsilon-\varepsilon_{
_L}}{\varepsilon_{_U}-\varepsilon_{_L}} }-\pi
\end{equation}
has been used. Above, $\varepsilon_{_U}$ and $\varepsilon_{_L}$ are
the upper and lower edges of the band gap, respectively.

\section{\label{results}Results and discussion}

\subsection {Electronic structure}\label{overview}

\begin{figure}[t]
\includegraphics[width=\columnwidth]{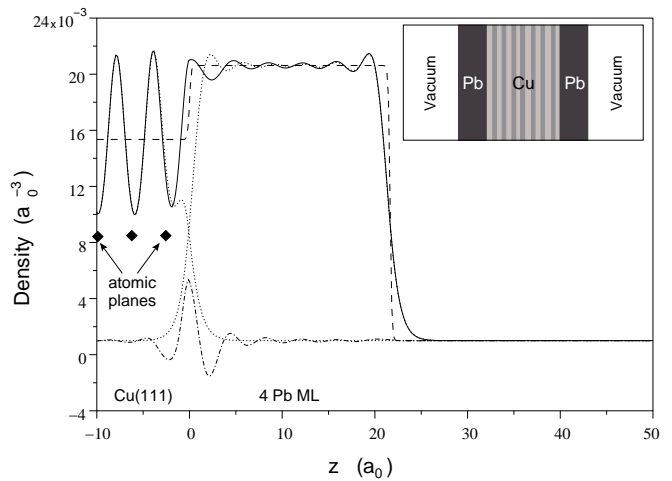}
\caption{\label{rho} Electron density (solid line) and average
positive background density (dashed line) for 4 ML's of Pb on the 
Cu(111). The origin $z=0$ is at the Cu-Pb interface.
The dotted curves show the electron densities of the
free-standing Cu(111) and Pb slabs. The dash-dotted
curve shows the charge transfer in the combined system relative
to the free-standing Cu(111) and Pb slabs
($\Delta \rho=\rho_{_{Pb/Cu(111)}}
\! -\rho_{_{Cu(111)}} \! -\rho_{_{Pb}}$). The inset sketchs the
symmetric geometry used in the calculations.}
\end{figure}

\begin{figure*}[t]
\includegraphics[width=2.\columnwidth]{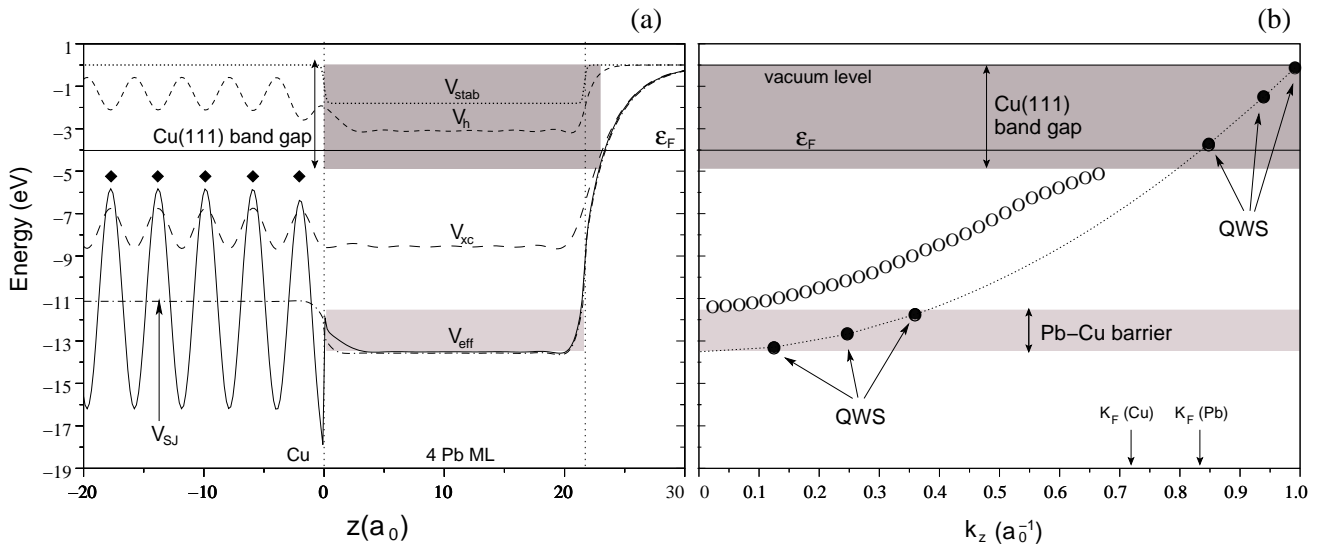}
\caption{\label{pot} (a) Effective potential 
$V_{eff}(z)$ (solid line) and its components, the
exchange-correlation potential $V_{xc}(z)$ (long-dashed line), Hartree
term $V_H(z)$ (short-dashed line) and Pb stabilization potential
$V_{stab}$ (dotted line) for 4 ML's of Pb on the 
Cu(111) surface. The origin $z=0$ is at the Cu-Pb interface
and the energies are measured with respect to the vacuum level.
The dash-dotted curve is the potential obtained by using the 
stabilized jellium 
model also for Cu. The straight horizontal line denotes the
Fermi level $\varepsilon_F$. The vertical dotted
lines represent the surfaces of the 4 ML Pb slab and the diamonds
mark the position of atomic Cu planes. The shadowed regions distinguish
the energy and space regions where localized QWS's can exist. 
(b) Energy eigenvalues as a function of the wavevector
$k_z$. The open circles represent
states extended over the whole system while the filled circles
correspond to QWS's trapped in the Pb layer. The dotted parabola is the
free-electron-model band in Pb plotted as a guide to the eye. The
dark gray and ligh gray areas correspond to the Cu(111) energy gap and
the potential well in Pb, respectively. The Cu and Pb
free-electron-model Fermi wavevectors are marked for reference.}
\end{figure*}

The main results of this paper are obtained with the pseudopotential
generated in the previous section for the Cu(111) surface and
the stabilized jellium model for the Pb overlayer. 
In the inset of Fig. \ref{rho} we sketch the geometry used in 
the calculations. In the middle there are 25 ML's of Cu
and Pb overlayers corresponding up to 23 ML's (1 ML of Pb is 5.41 $a_0$) 
are attached symmetrically on both sides.
Complementary calculations have been performed
for unsupported Pb slabs, too.

Fig. \ref{rho} shows the electron and positive background
densities corresponding to the coverage of 4 ML's of Pb. 
We notice in the electron density in Pb six Friedel oscillations
with the wavelength of about half Fermi wavelenght.
In the present case they are produced by the 6th QWS band, 
the bottom of which has dropped well below the Fermi level. 
The 7th QWS band is still unoccupied. This kind of well-developed
Friedel oscillation pattern commensurate with the thickness of
the overlayer is characteristic for the stable
``closed shell'' or ``magic'' overlayer systems.
In Cu(111) the electron density oscillates
strongly, according to Fig. \ref{rho}, as a consequence of the oscillating pseudopotential. 

Fig. \ref{rho} gives also the electron densities of the corresponding 
free-standing Cu(111) and Pb slabs as dotted
curves. It is remarkable that the electron density moves from Pb towards
Cu(111), indicating a possible increase in the effective
width of the Pb slab. The charge transfer obtained when comparing the
Pb/Cu(111) system with its free-standing counterparts ($\Delta
\rho=\rho_{_{Pb/Cu(111)}}\! -\rho_{_{Cu(111)}} \! - \rho_{_{Pb}}$)
reveals charge accumulation at the interface. 
A small amount of charge ($ \sim
5\% $ of one Pb ML charge) is transferred from Pb to Cu in order to
equilibrate their chemical potentials, $\sim-4.1$ eV and $-4.94$ eV
for Pb and Cu, respectively. 
The charge transfer oscillates slightly 
as a function of the Pb slab thickness due to the oscillations in the
Pb/Cu(111) Fermi level. These oscillations are in turn a consequence 
of the electron $z$-confinement in the Pb overlayer.

In Fig. \ref{pot}(a) the effective potential $V_{eff}$ and its
components $V_{xc}$, $V_H$ and $V_{stab}$ are shown; the total
pseudopotential $V_{ps}$ is not plotted for clearness.
The effective potential corresponding to the same system, but
modeled with the stabilized jellium for Cu, is also
drawn (dash-dotted line), for comparison. Notice that it gives
approximately the mean value of the Cu(111) pseudopotential. 
The dark gray shadowing gives the rough spatial
extension where the localized QWS's can appear because of the 
Cu(111) energy gap. 
The light gray area marks the potential well between the vacuum 
barrier and the Pb-Cu interface barrier. This well is created because
the potential in Pb is $\sim$ 2 eV lower than the average Cu
potential (notice the dash-dotted curve).

In Fig. \ref{pot}(b) the eigenvalues
are plotted as a function of the wavevector $k_z$. 
Actually, as will be explained below, it is not a straightforward 
task to assign a wavevector $k_z$ to each state. Again, the
dark gray area marks the position of the energy gap induced by the
Cu(111) potential and the light gray area shows the lowest lying QWS's
bound by the Pb-Cu interface barrier. The QWS's appearing in these energy
regions are plotted with filled circles (Fig. \ref{wave2} shows
an example of the corresponding eigenfunctions (solid line)).
The QWS's fall on the parabola (dotted line) which is the 
free-electron-model band for Pb. The states extending over the whole 
Cu-Pb system and forming a continuous band are plotted with open circles.
The nearly parabolic band reflects the free-electron character in 
our description of the Cu substrate. 
The QWS's in the lower, light gray area are supposed 
to have a minor relevance on the electronic properties of the system.
In contrast, QWS's in the upper dark gray area play an important role, 
because increasing the Pb thickness lowers the QWS energy and 
they become occupied one by one.

The repeated occupation of new QWS's as a function of the overlayer
thickness produces oscillations in all the electronic properties. In
particular, oscillations in the energy give rise to Pb overlayers
of especial stability. We studied recently these "magic" 
island heights using the present model for the electronic 
structures.\cite{edu_slab}


As pointed out above, we divide the eigenstates into QWS's and
bulk states. We use different methods to assign wavevectors 
$k_z$ for these two groups of states.
For bulk states, of minor importance in this study, we use the
\emph{particle-in-a-box} wavevector $k_z=\pi n/D$, where
$n$ is an integer and $D$ the box width. On the other hand,
QWS's are very localized in the  Pb slab and we obtain their
wavevectors from the confinement kinetic energy $\epsilon_k$
as $k_z=\sqrt{2\epsilon_k}$. $\epsilon_k$ is the energy difference between
the QWS energy eigenvalue and the average potential value
around the center of the Pb slab. The wavelengths can be
measured also from the eigenfunctions and similar values are obtained. 
Below, the $k_z$ vector turns out to be a relevant parameter when evaluating 
the phase shifts at the interfaces.

\subsection {Determination of confinement barriers I} \label{barrier}

\subsubsection{Comparison with the experiments}

Scanning tunneling spectroscopy is capable of resolving 
energy levels of QWS's. Nevertheless, the interpretation 
of the experimental results is still difficult.\cite{yeh2,otero2}
The results obtained by our model provide a good tool to 
enlighten the problems encountered and to suggest a simple but realistic 
picture with relevant parameters.

\begin{figure}
\includegraphics[width=\columnwidth]{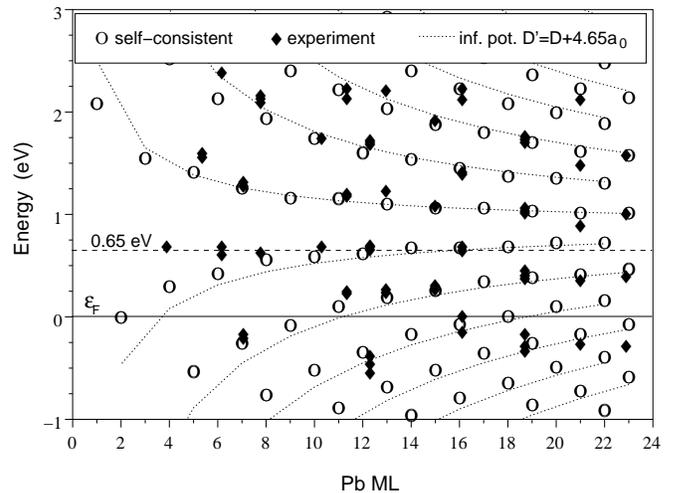}
\caption{\label{eigen} Energy eigenvalues of QWS's in the Pb
overlayer on the Cu(111) surface as a function of the number Pb ML's.
Open circles and filled diamonds give the calculated and 
experimental STS \cite{otero2} values, respectively.
The dotted curves link points calculated with the infinite potential 
well model (See, Sec. \ref{analy_for}) with the effective well width of
$D'=D+4.65a_0$. $D$ is the nominal width given by the number
of Pb ML's. The energies are given with respect to the Fermi level.}
\end{figure}

Fig. \ref{eigen} shows the calculated eigenvalues as a
function of the number of Pb ML's. The comparison with the QWS energy
levels measured by Otero {\it et al.}\cite{otero2} shows a good
agreement for large coverage heights.
Below 6 ML's the correspondence is a bit worse. This can be due to an
inaccurate description of the Cu-Pb interface or due to the
interaction with the Cu $d$ electrons omitted in the calculations.
The importance of these phenomena decreases as the number of ML's
increases.

To explain the measured energy eigenvalues Otero 
{\it et al.}\cite{otero2} tried first the finite-square-potential-well 
model but the fit was not satisfactory. However, they obtained 
a very good agreement by using the infinite-square-potential-well model 
or by decreasing the width of the finite square well by $\sim 2.6a_0$.
The fact that the infinite potential well produces much better
results than the, in principle, more realistic finite potential well is
counterintuitive, because it has been proven for
slabs,\cite{stratton,schulte} wires \cite{edu,garcia} and clusters
\cite{brack} that the real potential profiles are soft.
In fact, it can be seen in
Fig. \ref{pot}a that the self-consistent potential is soft  with
an effective width increasing as a function of the QWS energy level. 
On the other hand, this effect seems not to be counterbalanced by the
confining potential on the Cu-Pb interface. The QWS's penetrate 
inside the Cu(111) (see Fig. \ref{wave2}) and the electron density in
Fig. \ref{rho} moves towards Cu. Nevertheless, a very good
agreement with the experiments is obtained with both the 
infinite-square-potential-well model and soft self-consistent potentials. 
In addition, in Fig. \ref{eigen} we show the eigenvalues 
(linked by dotted curves) obtained with the 
infinite-square-potential-well model but by using well widths 
increased by $4.65a_0$ in order to take into account the 
wavefunction spill-out. As can be seen, a good fit is obtained
also in this way.

\subsubsection{Calculation of the phase shifts}

In order to clarify the reasons for these contradictions or dissimilar
results, we use the phase-accumulation model and evaluate the phase
shifts for the confinement barriers in our self-consistent
calculations. We use the unsupported Pb slab
to calculate the phase shift  $\phi_{Pb-vac}$ and then we can
evaluate the phase shift $\phi_{Cu(111)-Pb}$ in the Pb/Cu(111) system.

The procedure used is to choose a QWS with an energy
just below the vacuum level and then to identify the corresponding
quantum number $n$. After the wavevector $k_z$ is calculated from the
kinetic energy $\varepsilon=k_z^2/2$ as explained at the end of
Section \ref{overview}, the phase shift $\phi_{Pb-vac}$ is evaluated 
for the QWS energy level by Eq. (\ref{eche}). Increasing the slab
thickness, the energy eigenvalue decreases and we calculate
$\phi_{Pb-vac}(\varepsilon)$ as a function of the energy. The phase
shift depends on the potential profile of the surface. Therefore we 
check that it remains
unaltered for thicknesses over $\sim 2$ ML's. Once
$\phi_{Pb-vac}(\varepsilon)$ has been calculated by using the unsupported
Pb slab, the phase shift $\phi_{Cu(111)-Pb}(\varepsilon)$ can 
be evaluated in the same way using the results of the Pb/Cu(111) system.
Similarly to the above procedure for the phase shifts, we also determine,
according to Eq. (\ref{eche2}), the shifts in the
surface and interface positions $\delta_{Pb-vac}(\epsilon)$ and
$\delta_{Cu(111)-Pb}(\epsilon)$, respectively.

The identification of the quantum number $n$ is easy for 
states in free-standing Pb slabs. But it is difficult for the
QWS's in the Pb/Cu(111) system because, in
principle, it is not known how many states are hidden as resonances
in the region of delocalized states in Cu (the white area 
between the gray ones in Fig. \ref{pot}(b). 
Nevertheless, the local DOS integrated over the Pb overlayer 
reveals the number of the QWS resonances and, in addition,
we have the possibility to plot the eigenfunctions in order to
check the identification.

\begin{figure}
\includegraphics[bb= 0 0 507 798,width=\columnwidth]{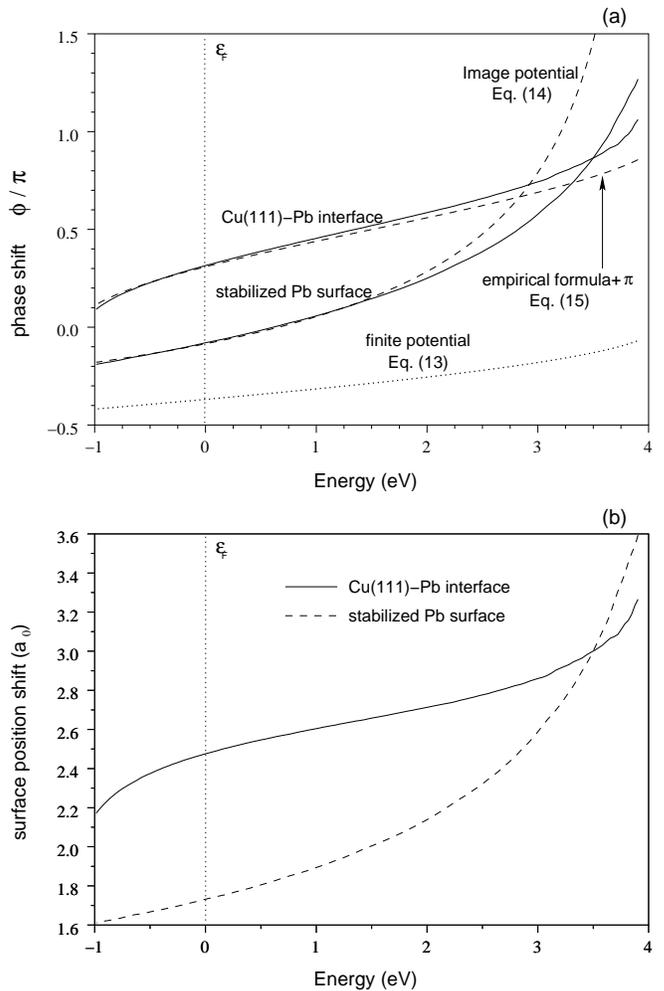}
\caption{\label{phase} (a)  Cu(111)-Pb and Pb-vacuum phase shifts as a
function of the QWS eigenenergy given with respect to the Fermi
level. The solid curves give the results
of our DFT calculations. The dashed curves denote the values
obtained by using Eqs. (\ref{imag_p}) and (\ref{empiric}) for the 
Pb-vacuum and Cu(111)-Pb phase shifts, respectively ($\pi$ is
added to Eq. (\ref{empiric})). The dotted curve corresponds to
the finite-potential-step model for the Pb-vacuum barrier. 
(b) Shifts in the effective Cu(111)-Pb interface and Pb surface positions
are given by solid and dashed curves, respectively.}
\end{figure}

The dependences of the phase shifts on the energy eigenvalue are shown in
Fig. \ref{phase}. We notice that nearly in the whole energy
range the phase shift at the Cu(111)-Pb interface is larger than that at
the vacuum side. This means that the wavefunction penetration into
Cu(111) is larger than the spill out into the vacuum. The behaviour 
of the phase shift $\phi_{Pb-vac}(\epsilon)$ can be compared with that
obtained in the three-dimensional DFT calculations with more 
realistic atomic pseudopotentials. For example, the 
results by  Wei and Chou\cite{wei} are in a qualitative agreement
with the preset ones. 

The phase shifts of the image-potential model, Eq. (\ref{imag_p}),
reproduce quite well our results with the exception of the energies 
close to the vacuum level. This is due to the exponential decay of
the DFT potential into the vacuum. The empirical Eq. (\ref{empiric}) 
gives good results at the Cu-Pb interface if one corrects for the known 
downward shift of $\pi$. \cite{pendry} The finite-square-potential-well model
does not produce phase shifts in agreement with the self-consistent results.

We plot in Fig. \ref{phase}(b) the shifts of the barriers in the
infinite-square-potential-well model according to Eq. (\ref{eche2})
as a function of the energy eigenvalue. In the energy window
from -1 to 3 eV accessible in the STS measurements the effective 
width increases more at the Cu(111)-Pb interface than at the 
Pb-vacuum surface. The shift of the effective Pb-vacuum barrier
is in agreement with the jellium-model calculations by 
Stratton\cite{stratton} who obtained the average shift of
$1.41a_0$ between the bottom of the potential well (at -9.5 eV)
and the Fermi level.

Combining the results of Fig. \ref{phase} and Eqs. (\ref{eche}) or
(\ref{eche2}), the eigenvalues in Fig. \ref{eigen} are reproduced.
(The energy eigenvalues have to be shifted downwards by the width
9.5 eV of the occupied energy band.) 
A good qualitative behaviour is obtained simply by using the mean
effective width $D'$ which is  the ideal width $D$ increased by
$\sim 4.65 a_0$. The dotted curves in Fig. \ref{eigen} link the
eigenenergies calculated in this way.

\begin{figure}
\includegraphics[width=\columnwidth]{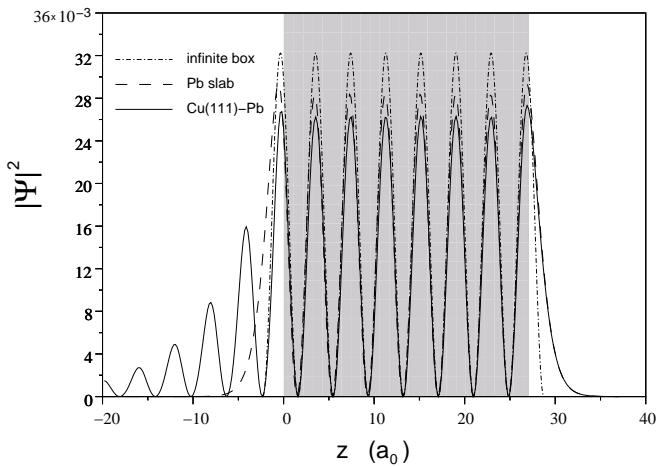}
\caption{\label{wave2} 
Electron density of a QWS in the 5 ML thick Pb overlayer on
Cu(111). The solid curve results from the self-consistent DFT
calculation for the Pb/Cu(111) system. The dashed and dash-dotted
curves are obtained using the free-standing Pb slab and
the infinite-square-potential-well model, respectively. 
The gray region denotes the Pb overlayer on Cu(111).
For the details, see the text.}
\end{figure}

We compare in Fig. \ref{wave2} our self-consistent DFT calculation
for the Pb/Cu(111) system with two simple model calculations.
There are 5 ML's of Pb on the Cu(111) surface and we plot the 
electron density of the QWS at $\varepsilon=0.53$ eV (solid line).
The corresponding state in a free-standing Pb slab (dashed line)
is obtained by shifting the left Pb-vacuum boundary by $0.75a_0$
to the left in order to mimic the larger penetration of the 
wavefunctions into Cu(111) than into vacuum. The value of $0.75a_0$
reproduces the correct infinite-barrier shift
in Fig. \ref{phase}(b). The eigenenergy of this Pb slab state is
also 0.53 eV. On the right surface, of course, both wavefunctions
overlap and at the left surface the slab wavefunction tries to
mimic, without oscillations, the decay of the wavefunction in the
Pb/Cu(111) system. The corresponding state calculated using the 
infinite-square-potential-well model with the appropriate shifts 
of the barriers
($1.66a_0$ and $2.37a_0$ from Fig. \ref{phase}(b)) is also given 
in Fig. \ref{wave2} (dash-dotted line). Note that this simple model
gives the correct energy eigenvalue and describes well the wavefunction 
inside Pb but it cannot account for the decay into the Cu substrate. 
In conclusion, the two simple models reproduce 
quite well the eigenfunctions inside Pb and at the Pb-vacuum
barrier, but not beyond the Cu(111)-Pb interface. In order to
describe properties related to the penetration of the wavefunctions
into the substrate, the slab and infinite-square-potential-well models 
are inadequate.

\subsubsection {Importance of quantum numbers}
The model based on the phase shifts or on the effective width increase of the 
infinite potential well reproduces the eigenvalues and eigenfunctions 
of the self-consistent calculation. Because the self-consistently 
calculated eigenvalues agree with the experimental ones, \cite{otero2}
the latter can be fitted with $\delta \approx 4 a_0$. Nevertheless, 
Otero {\it et al.} \cite{otero2} obtain a good agreement with $\delta=0$.

To explain the contradiction we have compared the 
infinite-square-potential-well 
energy spectrum (Eq. (\ref{eche2}) obtained with $\delta=0$) and our model
results corresponding to $\delta \not=0$. In the latter the
quantum number $n$ for the eigenstates with the nearly constant eigenenergy
of $0.65 eV$ are  $n=4,7,10,\ldots$ for 2,4,6,\ldots ML's of Pb, respectively.
Actually, this is in agreement with the pseudopotential calculations
for free-standing Pb slabs. \cite{wei} But $\delta=0$ gives the 
corresponding quantum numbers as $n=3,6,9,\ldots$ \emph{i.e.}, they 
are one unit smaller than the correct set. This explains why both models 
reproduce approximately the same experimental energy spectrum.
The wavelength of the states $n=3,6,9,\ldots$ at 0.65 eV in
the infinite potential well is $\lambda_{0.65}=7.2 a_0$.
When we increase the width of the well by $\lambda_{0.65}/2$ the
states $n=4,7,10,\ldots$ lie exactly at the same
energy. This does not hold for energies far from $0.65eV$ 
and the eigenenergy spectra become remarkably different. 
Nevertheless, in the narrow energy window from -1 to 3 eV scanned in
experiments, the energy spectra of both models are very
similar and fit quite well the experimental results. 
As a matter of fact, an even better fit of experiments seems to be obtained
by increasing the width of the potential well by $\lambda_{0.65}$ 
and using states with quantum numbers $n$ two units larger than
in the original infinite-square-potential-well model.

In conclusion, the energy spectra can be fitted with several different 
thickness and quantum numbers. The danger is that other properties 
such as the wavefunctions or "magic" heights are wrongly determined. 
For example, the correct determination of the physical parameters reveals 
crucial in the description of the envelope function which 
plays a key role in determining the energetics of the quantum well states 
\cite{kawakami} or the magnetic coupling in multilayer 
structures.\cite{ortega2}

\subsection {Determination of confinement barriers II}

\subsubsection{Calculation of $\delta$ from energy differences}

The disadvantage of the previous method for determining the
positions of the confining barriers is the necessity to measure
a large number of QWS eigenenergies in order to label correctly the quantum
states. However, there exists methods in which the knowlegde of the
quantum number $n$ is not needed. \cite{altfeder,su,su2,wei}
We now elaborate this kind of methods.

From the energy difference $\Delta_n$ between two consecutive states, one can
obtain an effective thickness $D'=D+\delta_0$ of the infinite potential well as
\begin{equation}\label{thick1}
\Delta_n=\varepsilon_{n+1}-\varepsilon_n=\frac{\pi^2(2n+1)}{2(D+\delta_0)^2}=
\frac{k_0\pi}{D+\delta_0}.
\end{equation}
Above, $D$ is the thickness of the positive background density of the
overlayer slab or, in the case of experiments, the 
measured thickness. $k_0=\frac{\pi (n+0.5)}{D+\delta_0}$ is 
a wavevector depending on $n$ and $D'$.
As before, $\delta_0$ is the
difference between the measured nominal value and the effective one
confining the electrons. The latter takes into account, e.g., the effect 
of the electron spill out into the vacuum or into the substrate, 
the unknown thickness of the wetting layer, and effects
due to stress or relaxation at the boundaries. 

We consider QWS's in overlayers of different thickness
and within a given window, e.g. the QWS's shown in Fig. \ref{eigen}.
We write Eq. (\ref {thick1}) as
\begin{equation}\label{thick2}
\Delta_n^{-1}=\frac{(D+\delta_0)}{k_0\pi}.
\end{equation}
Then, by plotting $\Delta_n^{-1}$ as a function of $D$ and by fitting a straight
line to the data, the slope gives 
$k_0$ averaged over the energy window in question. For example,
the theoretical data of Fig. \ref{eigen} gives $k_0=0.88a_0^{-1}$ 
which corresponds to the kinetic energy of $10.5eV$, i.e. an energy 
$\sim$1 eV above the Fermi level. 

\begin{figure}
\includegraphics[width=\columnwidth]{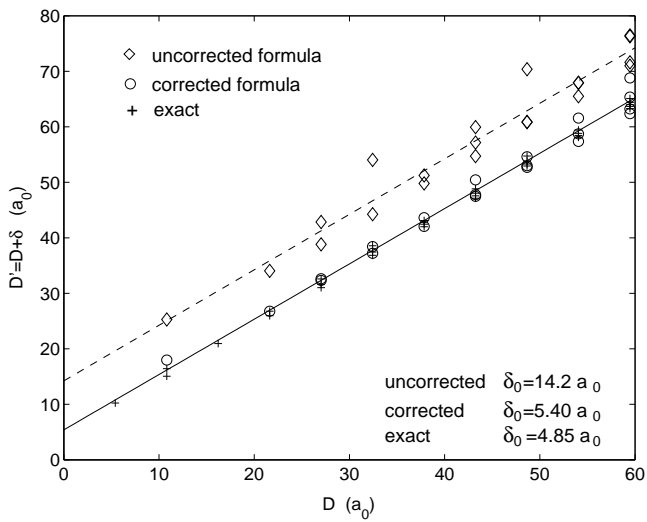}
\caption{\label{straight} Effective thickness $k_0 \pi \Delta^{-1}_n$ 
calculated from the theoretical QWS energies in Fig. \ref{eigen}
(energy window from -1 to 3.5 eV) as a function of the Pb 
jellium slab thickness. The diamonds and dashed line correspond to the data and fit according to Eq. (\ref{thick2}).
 The circles and the solid
straight line are the data and the fit according to Eq. (\ref{correc3}).
The crosses are the exact effective thickness calculated
from $D'=\pi n / \sqrt{2\varepsilon_n}$.
Data for 23 ML's are used in the fitting, but only data for 11 ML's are
plotted for clarity.}
\end{figure}

The original theoretical data and the fit are shown in Fig. 
\ref{straight} by the diamond markers and by the dashed straight line, 
respectively. 
We plot $D'=k_0 \pi \Delta_n^{-1}$, instead of $\Delta_n^{-1}$, as a 
function of $D$. This is because we want to include 
the exact values $D'=\pi n / \sqrt{2\varepsilon_n}$
and the results of Eq. (\ref{correc3}) (see below).
The intersection of the dashed straight line with the vertical axis 
gives $\delta_0=14a_0$. This value is much larger than the value 
obtained for the total shift of the
potential barriers in Section \ref{barrier}. Namely, at the energy of $1eV$
above the Fermi level we obtain from Fig. \ref{phase}(b) that
$\delta_0=4.5a_0$.

\begin{figure}
\includegraphics[width=\columnwidth]{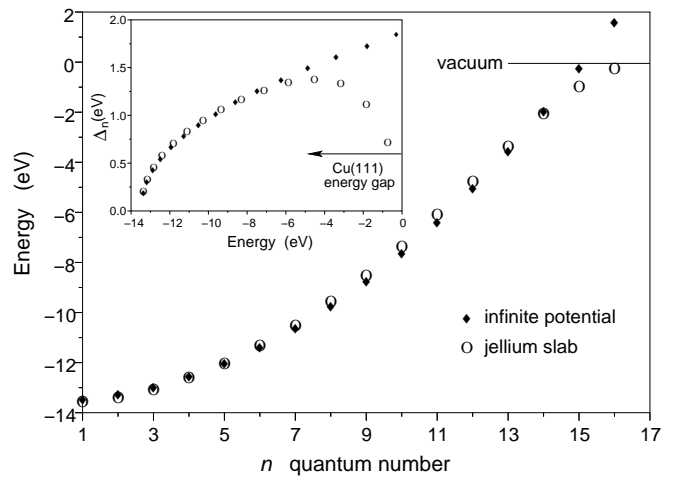}
\caption{\label{8MLPb} 
Energy eigenvalues (bottoms of subbands) of a free-standing
8 ML thick Pb slab as a function of the quantum number $n$.
The open circles and the filled diamonds give the results
of the stabilized-jellium and infinite-potential-well models, respectively.
In the latter, the width of the well corresponds to 8 ML widened
by 4.5 $a_0$ in order to take into account the electron spill out.
The vacuum level defines the zero energy level.
The inset shows the energy differences between the consecutive
states as a function of the energy. 
The energy range of the QWS's in the Pb/Cu(111) system is given. 
}
\end{figure}

The method described above (Eq. (\ref{thick2})) has been used to determine 
the number of Pb wetting layers in the STS. Our finding that the
method overestimates the distance between the confinement barriers
may explain the large values \cite{altfeder,su,su2,wei} obtained by 
this scheme in comparison with other experiments. \cite{yeh2,hupalo,mans}
The reason for this disagreement 
is the basic fact that Eqs. (\ref{thick1}) and (\ref{thick2}) 
correspond to the infinite potential well, whereas a finite
potential well is closer to reality. The infinite-square-potential-well 
model with the shifts $\delta_0$ of the confining barriers reproduces quite 
accurately the energy spectrum as is demonstrated in Fig. \ref{8MLPb}.
The figure shows that the infinite potential well with the thickness 
of 8ML+4.5$a_0$ mimics the energy spectrum of a 8 ML unsupported 
Pb jellium slab. But the reasoning to the opposite direction,
the determination of the $\delta_0$ value from the real energy spectrum
cannot be done using the scheme.
The explanation is that the energy eigenvalue differences 
$\Delta_n$ between the 
consecutive states behave differently in the infinite-square-potential-well
and in real systems. The inset of Fig. \ref{8MLPb} shows that
the infinite potential well results in a monotonically increasing
$\Delta_n$ whereas in the more realistic stabilized-jellium model
$\Delta_n$ decreases close to the vacuum level. The wrong trend
in the infinite-square-potential-well model is compensated by the erroneously 
large effective width of the Pb slab or overlayer obtained by
applying Eq. (\ref{thick2}). In the next subsections we suggest
a method purified from this effect.

\subsubsection{Corrected formula for the calculation of $\delta_0$}

The option we choose to correct the overestimation inherent in
Eq. (\ref{thick2}) is to introduce the effect of the finite 
potential barrier. This can be done by assuming that the surface shift 
is energy-dependent as $\delta=\delta_0+\delta(\varepsilon)$. Here,
$\delta_0$ is the mean value we want to determine. 
Our aim is to obtain information about the electron confinement 
strength through the 
$\delta_0$ parameter, which is energy-independent but it is
reproduces satisfactorily the energy spectrum (see Fig. \ref{8MLPb}). 
Nevertheless, it is necessary to use the energy-dependent 
$\delta(\varepsilon)$ function to obtain relevant $\delta_0$ values.

The energy difference between the successive states can be obtained 
to the first order as $\Delta_n=\frac{{\rm d}\varepsilon}{{\rm d}n}$ ($n$
increases by unity). The energy derivative for a given thickness 
$D'$ is
\begin{equation}\label{correc}
\frac {{\rm d} \varepsilon}{{\rm d}n}=\frac{\pi^2
n}{D'^2}- \frac{\pi^2n^2}{D'^3}\dot\delta(\varepsilon) 
\frac{ {\rm d}\varepsilon}{{\rm d} n}
\end{equation}
where $\dot\delta(\varepsilon)$ is the energy derivative of the 
surface shift and it depends on energy. This
equation has to be evaluated for a given $n$. For the infinitely 
deep potential well $\dot \delta=0$ and we do not recover the exact
result of Eq. (\ref{thick1}). 
This deficit can be corrected by evaluating the right-hand side of
Eq. (\ref{correc}) at $(n+1/2)$. The corresponding $\dot \delta$ 
is also evaluated at $(\varepsilon_{n+1}+\varepsilon_n)/2$
using the self-consistent $\delta$ values shown in Fig. (\ref{phase}b).
Then, rearranging the terms of Eq. (\ref{correc}) we obtain the corrected
formula for the thickness as
\begin{equation}\label{correc3}
\pi k_0 \Delta^{-1}_n-k_0^2\ \dot\delta=D'=D+\delta.
\end{equation} 
Here, it is accurate enough to use the $k_0$ obtained previously
with Eq. (\ref{thick1}). Omitting 
the dependence of $\delta$ on energy on the right hand side of 
Eq. (\ref{correc3}), i.e. $\delta=\delta_0$, we can calculate the
value of $\delta_0$ by fitting a straight line for $D'$ as a
function of $D$. 

The circles and the solid straight line in Fig. \ref{straight} give the 
$D'$ values and the fit, respectively, corresponding to 
Eq. (\ref{correc3}). The new mean value $\delta_0=5.4a_0$ is 
consistent with the electron spill out shown in Fig. (\ref{phase}b) and 
it is much better than the value of $14a_0$ obtained with the 
uncorrected formula (\ref{thick2}). Namely, the exact
values $D'=\pi n / \sqrt{2\varepsilon_n}$ (shown as crosses
in  Fig. \ref{phase}) give $\delta_0=4.85a_0$.

\subsubsection{Analytic models for the $k_0^2 \dot \delta$ correction}

In order to apply the corrected scheme of the previous subsection
the energy-dependent derivative $\dot\delta(\varepsilon)$ has to be known.
We study now the reliability of different analytical models in its
estimation. These models allow us to extend the previous 
analysis to other substrates or overlayers without doing 
self-consistent electronic structure calculations.

\begin{figure}
\includegraphics[width=\columnwidth]{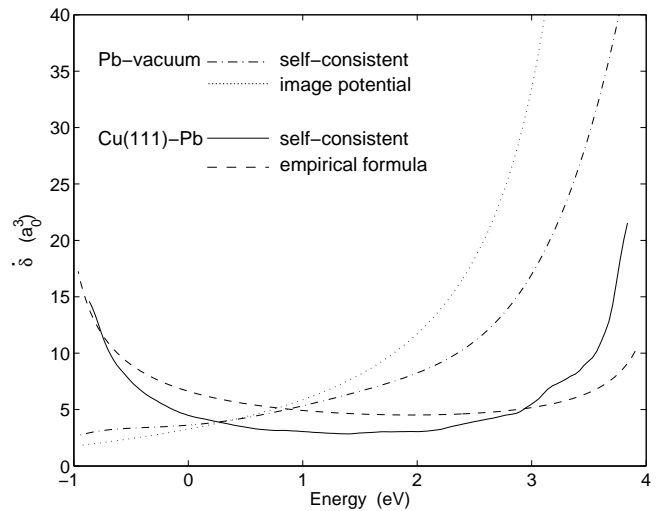}
\caption{\label{correction} Derivative $\dot \delta(\varepsilon)$
of the effective thickness with respect to the energy according to 
different models. The solid and dash-dotted curves give the self-consistent 
results for the interface and surface components, respectively. 
The dotted curve shows the analytical result for the surface 
component obtained by using the image potential model of Eq. 
(\ref{imag_p}). The dash-dotted curve gives the result for the interface
component obtained by using the empirical formula (\ref{empiric}). The energy origin is at the Fermi level.
}
\end{figure}

The finite-square-potential-well model does not provide a correction
large enough, because the potential is not a continuous
function of energy. The $\dot\delta$ values are too small (see 
Fig. \ref{phase}(a)).
The image-potential model (Eq. (10)) and the empiric phase 
shifts (Eq. (\ref{empiric})) provides results in a much better agreement with 
experiments.  We want to emphazise that even if
the analytical phase shifts may not produce the same $\delta_0$
values as the self-consistent calculations (for example, due to the 
fact that the phase shift is determined with the module of $2\pi$) 
they reproduce reasonably well the energy derivative $\dot\delta$ 
needed in the correction in Eq. (\ref{correc3}).

Fig. \ref{correction} shows the derivatives $\dot\delta$ from the
self-consistent calculation and from the analytical models of Sec.
\ref{analy_for}. The agreement is quite good, 
even if the derivatives of the self-consistent calculation
are generally smaller. Comparing self-consistent and analythical curves we notice that small differences in the phases (Fig. (\ref{phase})) produce big differences in Fig. \ref{correction}. As a matter of fact, at energies close to
the vacuum the image-potential results could be more relevant
than the self-consistent DFT results, which are affected by
the exponential decay of the DFT potential.
Because according to Fig. 8 the correction is large at energies
close to the vacuum level and also close to the bottom of the
projected energy gap of the substrate, we recommend the use of
QWS's at intermediate energies when determining the overlayer
thickness.

\begin{table*}[t]
\caption{\label{table_corr} 
$\delta_0$'s (in Bohr radii $a_0$) determining the effective widths of 
the infinite potential wells fitting the QWS energy spectra. 
The values are calculated from different 
theoretical and experimental spectra by using the uncorrected sheme 
of Eq. (\ref{thick2}) as well as the scheme of Eq. (\ref{correc3})) 
with the numerical correction obtained from the self-consistently calculated 
results or from the analytical image-potential and empiric formulae. 
The exact values are obtained by employing the equation 
$D'=\pi n / \sqrt{2\varepsilon_n}$ with the known $n$ quantum numbers.
The maximun overlayer thickness (in ML's) used in fitting are given 
in the second column. The numbers in parenthesis give the linear 
regression errors. The QWS of the DFT calculations used in fitting
span the energy range from $\sim 1 eV$ below Fermi level,
i.e. from the bottom of the Cu(111) band gap, to $\sim 1 eV$
below vacuum level (for Pb/Cu(111) corresponds to the experimentaly measured energy range of QWS's).}
\begin{ruledtabular}
\begin{tabular}{l c c c c c}
System &  Up to & $\delta_0$ from Eq. (\ref{thick2}) & $\delta_0$ from
 Eq. (\ref{correc3}) & $\delta_0$ from Eq. (\ref{correc3}) & Exact
 $\delta_0$ \\
 &  & uncorrected & numerical correction & analytical 
 correction & \\
 \hline
 Free-standing Pb  & 10 ML & 12(3) & 5.5(0.6) & 6(3) & 3.9(0.3) \\
 Pb/Cu(111) & 23ML & 12(1) & 5(2) & 3(2) & 4.7(0.1) \\
 Pb/Cu(111) & 10ML & 13(1) & 6(1) & 4(2) & 4.8(0.3) \\
 Experiments \cite{otero2} & 24ML & 15(5) & 7(5) & 5(5)  & 4.0(0.2)\\
 Na/Cu(111) & 20ML & 7(3) & 1(3) & 3(4) & 1.2(0.3) \\
\end{tabular}
\end{ruledtabular}
\end{table*}

Table \ref{table_corr} shows a collection of $\delta_0$ values
obtained for several systems using different approximations.
In addition to our theoretical results for the Pb/Cu(111) and the
free-standing Pb slab systems we analyze our similar results for Na/Cu(111)
and the experimental QWS spectrum of Pb/Cu(111) [\onlinecite{otero2}].
The "exact $\delta_0$" values are obtained by fitting a straight 
line through the exact results $D'=\pi n / \sqrt{2 \varepsilon_n}$.
In general, the uncorrected Eq. (\ref{thick2}) gives for the Pb systems
$\delta_0$ values nearly three times larger than the exact one. 
 In contrast, the corrected value offers a sub-monolayer
accuracy. 

For the Na/Cu(111) system the difference between the corrected and uncorrected $\delta_0$ is large as well. Although the wavelength of Na is larger than for Pb, $\delta_0$ does not scale with it. The deeper potential of Cu(111) than the Na potential (the opposite to the Pb/Cu(111) system) shifts the barrier inside the Na, i.e. the value of $\delta_{Cu(111)-Na}$ is negative. This explains the small value of $\delta_0$.

It is noticeable that applying Eq. (\ref{correc3}) with
the numerical and the analytical corrections to the experimental
results of ref. [\onlinecite{otero2}] we obtain $\delta_0=7.3a_0$ and
$4.7a_0$, respectively. These values are much smaller than the 
uncorrected value of $14.9a_0$. When we compare with the exact 
 value of $4.0a_0$, which is obtained by identifying first 
the $n$ quantum numbers. The linear regression error is anyway substantial. 
In addition to the errors due to our first-order approximation
in Eq. (14), the experimental dispersion in the QWS spectrum cause
uncertainty. In general, for self-consistent calculations the numerical correction provides better determination of $\delta_0$ compared to the analytical one. But in the case of the experimental data the analytical correction is better.

When analyzing experiments, the $\delta_0$ value obtained by the 
corrected scheme of Eq. (15) can be used as an initial parameter 
to determine the $n$ quantum number. Then,
employing the exact equation to fit the energy spectrum,
improved results are obtained.

\section{\label{conclusions}Conclusions}

We have performed self-consistent DFT calculations
to study the confinement barriers of electrons in Pb 
islands grown on the Cu(111) substrate. Calculations have been done for
free-standing Pb slabs and Pb slabs on Cu(111). Pb was described
by stabilized jellium and the Cu(111) substrate by a 1D-pseudopotential.
The model reproduces the most important physical properties and 
gives results in a good agreement with experiments.

The energies and wavefunctions of the quantum well states
in the Pb slabs characterize the confinement barriers at the Pb-vacuum 
surface and at the Cu(111)-Pb interface. We have analyzed these
states by using the phase accumulation model and by determining
effective widths of infinite potential wells reproducing the 
energies. The Pb-vacuum phase shift is in a
good agreement with more realistic pseudopotential calculations. 
The Cu(111)-Pb phase shift or the effective width of the potential well
accounts for the confining strength of the Cu(111) energy gap.
This strength is weaker than that of the Pb-vacuum barrier.

The information provided by our calculations and analysis allows
to improve the interpretation of QWS spectra measured by scanning 
tunneling spectroscopy.
More specifically, we have shown that a formula commonly used 
in the literature results easily in the overestimation of the effective
width of the infinite potential well. We have offered an alternative expression
to correct that deficiency, specially important for electronic high density (small $\lambda_F$) metals. The obtained results
can be used to estimate the width of the potential well and to
determine the quantum numbers for a more accurate analysis
of the confinement barriers.

Finally, the 1D-pseudopotential scheme provides a method for future studies of nanostructures on solid surfaces. Despite this work is focused on STS experiments and Pb/Cu(111) system, our results can be qualitatively applied to metallic nanoislands on semiconducting substrates, e.g. Pb or Ag on Si, as well as to photoemission experiments.

\begin{acknowledgments}
The authors are grateful to R. Miranda for his valuable comments. We acknowledge partial support by the University of the Basque
Country (9/UPV00224.310-14553/2002), the Basque Hezkuntza Unibertsitate eta Ikerkuntza Saila, and Spanish Ministerio de Ciencia y Tecnolog\'ia (MAT 2001-0946 and MAT2002-04087-CO2-O1).
This work was also partially supported by the Academy of Finland through 
its Centre of Excellence Program (2000-2005).
\end{acknowledgments}


\bibliography{bibliografia}

\end{document}